# PaaS Cloud — The Business Perspective

## Business Prospective of Platform-as-a-Service Cloud Computing

**Robail Yasrab** (robail@mail.ustc.edu.cn)[1,]

Department of Computer Science,
University of Science and Technology of China,
Hefei, China.

*Abstract*— The next generation of PaaS technology accomplishes the true promise of object-oriented and 4GLs development with less effort. Now PaaS is becoming one of the core technical services for application development organizations. PaaS offers a resourceful and agile approach to develop, operate and deploy applications in a cost-effective manner. It is now turning out to be one of the preferred choices throughout the world, especially for globally distributed development environment. However it still lacks the scale of popularity and acceptance which Software-as-a-Service (SaaS) and Infrastructure-as-a-Service (IaaS) have attained. PaaS offers a promising future with novel technology architecture and evolutionary development approach. In this article, we identify the strengths, weaknesses, opportunities and threats for the PaaS industry. We then identify the various issues that will affect the different stakeholders of PaaS industry. This research will outline a set of recommendations for the PaaS practitioners to better manage this technology. For PaaS technology researchers, we also outline the number of research areas that need attention in coming future. Finally, we also included an online survey to outline PaaS technology market leaders. This will facilitate PaaS technology practitioners to have a more deep insight into market trends and technologies.

I. INTRODUCTION

In the current era information technology plays a key role to excel business processes. That leads to attain a better competitive edge in the market. The state-of-the-art IT technologies are being in practice in most of groomed organizations [1]. Currently "Cloud Computing", emerged as a novel computing model available through the Internet [2]. It offers organizations with several innovative IT-based tools and services that benefit to businesses, such as agility, reasonable cost and greater efficiency [3]. It has been proved that application of the cloud computing improves the organizational processes and productive. According to [4], cloud computing technologies SaaS, IaaS and PaaS are transforming the traditional ways of computing. SaaS and IaaS are well-established areas of cloud computing, while PaaS is now emerging as one of the powerful technology paradigms [4].

PaaS cloud services are helping developers and businesses to accelerate innovation, enhance productive and deliver more business value in a much faster way. Currently, organizations are buying "as a service" cloud resources like applications (SaaS), infrastructure (IaaS) and platform (PaaS). PaaS is one of exceptional cloud technology because it offers tools, services and templates to develop powerful applications at the expense of low cost and time. Presently, we can have whole application development environment like coding, testing and delivery services available through online PaaS services provider.

PaaS transformed the traditional practices of the application designing, development, testing and deployment. PaaS offers all capabilities to incorporate innovations, flexibilities and low cost that cloud grows a business to the top line. A comparison between traditional and novel PaaS software development lifecycles has been shown in Fig. 1. Now PaaS is emerging as an innovative tool for organization to accelerate innovation in software development and delivering to the market in a faster way [5].

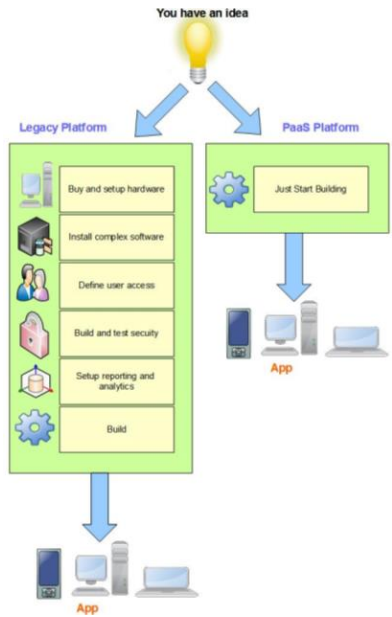

Figure 1 PaaS vs. Traditional Development Approach

According to [6], the current PaaS technology consumers (developers) don't need to be bothered with managing or controlling the underlying application development infrastructure (software packages, servers, network, operating systems, or storage). Aforementioned capabilities provided, managed and supported by the cloud service providers. [7] stated that, PaaS conceals the complexity of logic running among client and service provider's virtual server.

PaaS providers are trying to make application development experience familiar to application developers. By offering common programming languages support (Java, C, Python and PHP), and through drag & drop business-logic apps. These apps lessen the programming effort required for implementing blocks of code [4].

PaaS technology providers offer variety of technology services with different delivery methods. Different PaaS platforms offering different styles of development such as Bungee Connect offers support to developers to build applications in online mode only. While others, like Google App Engine (GAE) allows developer to develop offline and then update on GAE server so that it can be hosted. Google App Engine achieves such services through creating an instance of the online server environment locally [8].

This research focuses on the PaaS technology business prospective. This article based on different sections. The first section defined introduction and key benefits of PaaS technology. The second section described the primary technological concepts, layered architecture and delivery patterns. The third section explained the key Players of PaaS market, their overview and analyses of the most powerful player of PaaS technology. The fourth section is one of the key sections in this research article that based on SWOT analysis of PaaS technology from business and developer observations. It will scrutinize possible strength and opportunities that PaaS technology currently offering. It will also point-out possible weaknesses and threats that PaaS is facing. The fifth section demonstrated the key stakeholders of PaaS technology, their roles and contributions. A PaaS application development life cycle is completely different from the traditional software development lifecycle which is described in sixth section. The seventh section is about PaaS technology revenue models. Those are novel and innovative to run a business and generate profit. The author recommends that PaaS technology is more reliable and could be used in any organizations to boost performances and fully described in recommendations section. The second last section addressed the PaaS's future dimensions and future market growth, expected in PaaS technology in future. The final section is about conclusion and references.

## II. KEY BENEFITS OF PAAS

PaaS has transformed the traditional ways of application development. The PaaS technology is promoted as a benefit to application development; it would be helpful to boost efficiency of development teams, improve process, fast development (especially mobile) and enhanced IT-business collaboration. It provides greater flexibility, speed and agility to the whole application development process. Moreover, it offers a heterogeneous and predictable application infrastructure. Now any organizations no need to bother with supplementary supporting applications.

By adopting PaaS technology, organizations become independent from maintaining separate infrastructure for application development tasks such as designing, configuring, building, testing and deploying. Further, they only need to buy PaaS services and configure according to the needs of every programmer. This improves efficiency and reduces risk through a simplified development process.

Besides, improving application development lifecycle, PaaS also offer great benefit to other supporting activities associated with development lifecycle. PaaS improves productivity; minimize software's cost and permits firms to release their products faster by hosting the whole application development environment online. PaaS technology offers support for developers to resolve issues regarding configuring servers for developing applications. Moreover, it offers scalable deployment environment as well as integration and implementation tools. It also helps developers to manage their storage subsystems, operating system, security or server patches. PaaS technology resolve customer issues regarding network interfaces (trying to acquire web services or tools to communicate to one another). PaaS technology providers deal aforementioned aspects and issues, so that application developers only focus on developing software and quality of the system [9].

There is a provision in PaaS technology to offer ability to programmer to utilize high capacity shared architecture. It provides the whole architecture in a simplified form that could be shared with other application developers to code and test application. These platforms facilitate for rapid and easy work group expansion as per requirement. The provision of "online code management" is the characteristic feature of PaaS architecture that facilitates developers to modify, update and access code remotely [8].

Facilitating skilled programmers is a very prominent feature of PaaS. Even, it also helps less skilled developers to develop apps. It offers less skilled developers with state-of-the-art development tools and middleware. PaaS tools do not need incredible development skills because it offers great assistance to the overall development process.

IaaS (Infrastructure-as-a-Service) platform facilitates its clients to approach technology resources, according to their needs. PaaS technology shows resemblance in some aspects to IaaS technology in terms of offering access to resources according to needs and capacities of developers. Now, there is no need of having many different independent tools [10]. Currently, application developers do not require configuring resource requirements. PaaS platforms automatically optimize and adjust the allocation of resources as per user's needs.

'Resource configuration' requires tools and additional cost. However, PaaS automated resource configuration process successfully. Those organizations in which PaaS technology is not in-practice, they spend more budget on software management, change and updates. PaaS technology providers offer flexible cost models (pay-per-user and pay-per-month) for clients. By accruing these models any organization can minimize their software licensing and yearly maintenance costs [2]. PaaS technology also offers "usage-based pricing model" as per usage of resources by developers. This model makes PaaS technology more valuable for application development as compared to traditional application development approaches [9, 10].

In addition to software services support, PaaS also offers hardware resources. In traditional application development approaches, server and storage needs are one of the key business overhead. Traditional development approaches require huge amounts of processing and storage space for developing, testing and deploying new applications. That is a computationally exhaustive task. Through PaaS technology platforms, businesses don't need to have surplus resources in reserve. PaaS technology comprises essential tools to be required for whole app development life cycle for instance development packages, data storage and server capabilities. This approach leads to low initial investment and a great deal of cost savings for a newly emerging business [2].

The process of application development and deployment puts a huge stress on network bandwidth within a data center. Developers and testers need to perform workload tests to evaluate systems efficiency and behavior under changing circumstances. Aforementioned situations often slow down the operation of other working applications. Otherwise, it requires to more bandwidth capacity. In these circumstances, PaaS offers testing to be performed in vendor' cloud, rather than in the personal data center. This resolves the network bandwidth problem.

Security is one of the main concerns in cloud infrastructure [11]. Currently, most of PaaS vendors are having much better and improved security management infrastructure. For example, Google App Engine allows administrator, to create a new project and invite developers to participate in development. Now developers can only write code, check the error logs, browse storage, and upload latest version. However, for ensuring high security, GAE developer's operations are limited. Developer cannot directly change operating system functionalities. GAE does not allow developers to take either too much CPU time or capture hardware or abuse other resources [12]. By these approaches, PaaS vendors are ensuring much better security for their clients.

### III. PRIMARY TECHNOLOGICAL CONCEPTS AND TERMINOLOGY

#### A. Basic Technology

According to definition, PaaS is offered as a service. It is accessible through the web with no need to ever install, manage, upgrade or host tools and applications. It is based on ad-hoc and on-demand cloud services that support fundamental characteristics of cloud computing. PaaS is flexible technology because it can be scaled up and down quickly as per client requirements. It offers a huge shared pool of computing tools and resources to deal surges. Business and developers can deploy their products in SaaS-way that consumers pay for what they use. PaaS is significantly different from traditional web platforms. Traditional web platforms need uploads, installations, downloads and hosting. While the idea of "As a Service" states that developers are having provision and can manage instances and services of the "platform on demand". A lot of benefits can be attained if applications are developed over PaaS. In this way, the developed apps will be fundamentally "cloud ready," taking full benefits of basic "as-a-service" infrastructure and cloud services delivery models.

Sometimes, misconception persist that PaaS is similar as other application frameworks, for instance .Net or Ruby on Rails. However, there is a negligible similarities persist among these architectures. There is no need of configuring, uploading, troubleshooting and licensing in PaaS architecture because it is offered through cloud service. PaaS and application frameworks may coexist to facilitate SaaS application and technology solutions, for example, SaaS Maker and Heroku platform. Aforementioned cloud based technology platforms offer deployment and integration of applications those were developed in a variety of programming languages.

#### B. Paas Platform Types

Currently, PaaS platforms are implemented in different styles. PaaS Service providers acquire different approaches to offer better services. Some PaaS architectures are linked to a specific operating-system while others are intrinsically tied to a specific environment. This section illustrates different types of PaaS technology services.

SaaS Environment Anchored PaaS: Nowadays, several cloud vendors offer SaaS as a core business services for their customers. To grow and ramify their business capabilities, some of SaaS vendors have developed ecosystems. Due to technological ecosystems, independent software vendors are able to run applications on top of the SaaS vendor's platform. PaaS technology architecture allows the independent software vendors to implement applications in the vendor's ecosystem. Here, few examples have been quoted of vendor platforms offering PaaS services anchored to a SaaS environment, for example Force.com, Workday, Google App Engine, AppScale and Intuit Developer Network [9].

Operating Environment Anchored PaaS: The next category of PaaS technology services where PaaS is tied to an operation environment. Such operation environment makes it easier to operate and perform specific actions within given environment. In this category, IaaS technology providers have initiated new offerings up to the software stack. Above mentioned PaaS vendors offer the entire nuts and bolts (the OS, networking, storage, etc.) as well as tools for application development to deployment. Those companies who offer such PaaS services comprise AWS Elastic Beanstalk, Windows Azure, IBM SmartCloud Application Services and AT&T PaaS. PaaS providers like the Amazon and Microsoft are offering their abstractions and APIs so that developers are able to develop or deploy applications through this support [9, 10].

Open PaaS: The aim of Open-platform PaaS is to encourage an open process and environment which are not link to a particular cloud computing implementation. Open PaaS providers permit developers to bring their own platform to the cloud, which leads to flexibility. However, it can add more cost and complexity. Such open platforms are suitable for a hybrid cloud environment as they permit deployment on both private and public clouds. Therefore, open PaaS ease the migration between cloud platforms. Popular Open PaaS vendors include Cloud Foundry, OpenShift, Engine Yard, CloudBees, OrangeScape, Apprenda, DotCloud, and CumuLogic [9].

*C. Paas Implimention Requirements*

The fundamental idea of PaaS technology is to offer better productivity for developer rather than managing computer systems, network and storage. PaaS offers productivity through abstracting out complexities during application development and implementation [13]. Such abstractions could be achieved through very comprehensive technology architecture. This section has a look into some of the key required components to implement PaaS architecture. These components may also be considered as major requirements to implement PaaS architecture.

- Support for multiple languages
- Support for multiple runtimes
- Application isolation at storage, compute and network level
- Virtual Machines (VMs) (VMs are best to offer application isolation, however present a performance overhead due to translation from Guest ISA to Host ISA. Another state-of-the-art alternative is the container based virtualization. Container technology model is replacing VMs and offering light weight virtualization.)
- CPU time fair sharing
- Application Lifecycle event management
- Easy management of application states

*D. PaaS Layered Architecture*

Attaining high performance, abstracting complex details, fast development, easy app management and lots of other capabilities can be achieved through PaaS. PaaS is based on very comprehensive and sophisticated cloud architecture. This architecture is the collection of independent technology layers. This section explains key layered architecture and specific components of PaaS layers.

In 3-layer cloud infrastructure, PaaS tier is the middle layer, contains a number of most signification elements of the overall cloud infrastructure. Fig. 2 shows a complete three layer architecture of cloud computing. PaaS layer clients are application developers who develop their software and deploy them on cloud platforms. PaaS offers to developers with a programming language level atmosphere comprises a set of tools and well-defined APIs that enable developers to fabricate interface among cloud applications and development environments. This whole arrangement of tools and services also speeds up the cloud applications deployment. PaaS offering to the developers numerous advantages like load balancing and automatic scaling to integration and other additional services for example email services, authentication services and user interface. Now application developers have the capability to incorporate other technology services to their software on-demand. This makes system development more easy, speed-up overall deployment time, and reduces the bugs. Through PaaS, characteristics of the operating system could be updated frequently. PaaS also offers capability for globally dispersed development teams to develop jointly on projects [12].

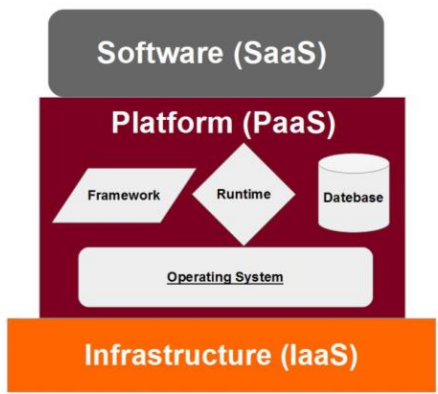

Figure 2 Three Layer architecture of Cloud Computing

Inside PaaS infrastructure, a key layer is "Database Management Systems (DBMs)" frequently act as a courier among the upper, lower and middle layers. This layer frequently comprises sophisticated security features and protocols. It also includes bots for organizing services and maintenance.

The second layer is "application server's layer". It offers correspondence among all activities in a specific system. In a number of cloud computing networks, whole hardware and server setup can be reserved for exclusive use of applications. Incorporating this layer in PaaS offers better results, improved technology optimization and fewer errors.

The most important layer in PaaS architecture is "Business Process Management Suites". It is a complex set of systems or applications that are assembled mutually in one arrangement. These sets are analyzed as per stakeholder requirements and then employed to deliver fuel for more application intensive processes. Nowadays, most of PaaS technology providers frequently use them to automate core business procedures.

Another significant layer is "Applications/Data integration" layer, where different elements and applications are responsible for data integration.

The outer layer is "Portals", which acts an access point among a user and cloud network. This channel located in the middle-ware of the PaaS network stack, which is used to link cloud resources [12].

*E. Generations of PaaS*

In order to improve PaaS technology, it is going through many transformations phases since it has been launched in the market. Currently, PaaS evolved to a great extent that business and developer start relying and adopting it. [13] has named this technology transformations in PaaS architecture as "PaaS generations".

Generation 01: This generation was based on classical fixed proprietary cloud platforms. AWS, Heroku and Azure were initial technology platforms who initiated such services.

Generation 02: This generation platforms were developed around open source solutions. OpenShift and Cloud Foundry were emerged as one of the top players of this generation. These technology providers offered clients to run their own PaaS (in the cloud or on-premise). Aforementioned, PaaS platforms also initiated the idea of containers and developed their own container models. However, currently these PaaS vendors are transforming their approach and moving toward improved technology models. For example: presently, Openshift transferred from its own container model to the Docker based container model. The same transformation performed by Cloud Foundry, through its internal Diego solution.

Generation 03: Currently, third generation of PaaS is purely focused on container models. The new PaaS platforms like Deis, Dawn, Octohost, Flynn and Tsuru are purely built on Docker based container models. These platforms build around Docker from scratch. Moreover, these PaaS models are deployable on public IaaS clouds or on their own servers.

*F. PaaS Technology and Industry Solutions*

The usage of PaaS varies within industry by software setups, businesses, businesses modes and corporate technology architectures. The integration designs among PaaS and industry solutions also differ accordingly. This underneath section elaborates such diversities among PaaS technology and industry solutions.

PaaS on industry solution: According to [14], this kind of PaaS cloud design functions for businesses which aim to host value-added cloud services. The entire system could work without PaaS technology because its central operations and architecture sustained through non-cloud infrastructure. The PaaS supported value-added constituents are test-beds for future commercial growth. For example, cloud supported "Self-Service Telecommunications Service" system that facilitates the end users for quick response.

PaaS in industry solution: [14] stated that this kind of framework executes inside a business solution through an embedded PaaS. For example, PaaS becomes a fragment of business solutions. One example that has been quoted is cloud supported integrated information frameworks which plays a key role in the petroleum and chemical industry. In this type of architecture, few functionalities or components are operational by PaaS technology services; however rest of the business setup is not hosted through the cloud.

PaaS of industry solution: In this model, both the industry solutions and PaaS are interconnected in the similar cloud infrastructure. This technology fusion is totally an industry specific PaaS. A usual instance of such kind of infrastructure is cloud supported "Financial Market data solutions". The objectives of such technology solution are to offer a customizable financial market data hub supported by PaaS. The PaaS allows an entire enterprise business model as a self-governing node in the industry network [14].

## IV. KEY PLAYER IN PaaS MARKET

Currently, there are numerous PaaS service providers in the market. Since to then, more and more companies and businesses are joining this cloud paradigm. Here is a list of some elite PaaS vendors, who are going toward improved services and better technology infrastructures.

Amazon Web Services (AWS): A major market leader in PaaS services. Amazon Web Services emerged primarily as an IaaS, however, later on also stepped into the PaaS through its AWS platform. AWS supports numerous languages like Python, Java, Perl, Ruby and other languages. AWS offer other services like Amazon Elastic Beanstalk for auto-scaling, application health monitoring and automatic load balancing, which are much supportive and helpful for application developers [15].

Heroku: Heroku is one of the initial cloud platforms, founded in 2007. Salesforce.com acquired the company in 2010, though it is still working as an independent subsidiary. It supports languages like Python, Ruby, Scala, Java, Node.js and Cloture. It based on the abstract computing infrastructure known as dynos, which runs processes in an isolated environment, based on virtualized Unix-style containers [16]. Heroku perform very efficiently with apps that support "Twelve-Factor-App" methodology. Now, Heroku is becoming more and more mature in service provision. This PaaS platform with numerous better capabilities (support third party apps e.g. Addons) is expected to become a major market shareholder of PaaS in future [17].

Google (Google App Engine): Google's App Engine (or GAE) is one of the initial PaaS market players, offering great PaaS services to-date. Google initiated its services in 2008 and incorporated most of the popular programming languages as well as frameworks for more flexible and user friendly application development. GAE empowers programmers to code locally (at host platform) through language-specific SDK and update it with other aware technologies like C++, JavaScript and Node. Google has improved its services extensively by the additions of recent enterprise competences like: availability of Service Level Agreements (SLAs), least planned downtime (99.95 percent), project lifecycle management, unified version control, sophisticated enterprise level support. By having all these capabilities, Google is guaranteed to dominate PaaS technology market in coming future [18, 19].

Microsoft Azure: Microsoft Azure launched in 2010, but shortly it turned out to a huge market competitor. Currently, Microsoft Azure offers wide-ranging cloud offerings. Businesses are able to bring their own technology infrastructure to the Azure cloud and utilize it to host their VMs and software. Different from other market leaders, Microsoft Azure is presenting a public-private cloud offering. That makes the user data more secure with greater privacy of significant intellectual property. To deal with the dynamic requirements of cloud computing, Window Azure offers a comprehensive service OS, service management and web hosting. With great language flexibility, Azure is striving to become a market giant [20, 21].

Salesforce.com (Force.com): Force.com is the PaaS division of the top cloud computing enterprise "Salesforce.com". It is recognized for establishing a standardized architecture of multi-tenancy. It offers capability to clients for application development by using a single instance of software. Force.com offers its clients; a built-in enterprise collaboration capability intended for application development. It also offers support for the mobile devices, so now a user can manage account through mobile app. According to website data, more than 3 million custom applications have already been developed on Salesforce PaaS cloud platform. This statistics indicate a bright and flourishing future of Salesforce business in coming years [22].

Red Hat (OpenShift): Red Hat OpenShift offer developers an access to advance, cost-effective, open and multilingual platform. Where developers can develop, test and deploy custom apps and implement across hybrid architecture. Through offering such technology features, Red Hat OpenShift is undoubtedly, the only open source PaaS vendor, offering complete and robust PaaS technology services [23]. OpenShift also offers management tools and apps for development procedures and their integration with other software systems. Moreover, it offers auto-scaling inside a multi-tenancy infrastructure. OpenShift offer clients with IaaS like characteristics, where developers are capable to control the hardware resource around their system. This is a big-plus for OpenShift as compared to other PaaS vendors.

Engine-Yard is one of the early cloud service providers. Now, it is getting pace, one of the main reasons is its strategic alliance with Microsoft in 2013. Through this alliance, developers will be able to use its open source PaaS capabilities, running on Microsoft cloud infrastructure. Engine-Yard is offering support to languages, like PHP, Ruby on Rails and Node.js. It also offers a great deal of capability for operations management, snapshot management, backup, cluster handling, load balancing and administering the database [24, 25]. Company is trying to be more and more vibrant in the open source cloud market by offering lots of innovative services and launching and collaborative projects with the world's well-known technology companies. As mentioned above, the strategic alliance of Microsoft and Engine Yard will deliver commercial grade application in the future.

Cloud Foundry was initially developed via VMware. Cloud Foundry obtains source code from Ruby users and developers. It is an open source PaaS platform that permits deployment of apps to Amazon Web Services (AWS), OpenStack, vCloud Air, vSphere and vCloud director. The key hosted services offered by Cloud Foundry are MongoDB, MySQL and RabbitMQ. Cloud Foundry's PaaS developers get a great deal of support and ease by having tools like Eclipse Plugin, command line, application scale and build integration. Currently, there is strong competition among Cloud Foundry, Heroku, AppScale and OpenShift [26]. On Dec 9, 2014 "Linux Foundation Collaborative Project" acquired Cloud Foundry as a collaborative project, so in future company is supposed to be a major competitor for the other open source PaaS companies. Cloud Foundry was initiated as open source PaaS vendor under the Apache 2.0 license. In February 2014, VMware declared the establishment of the Cloud Foundry Foundation and granted Platinum sponsorship to EMC, Pivotal [27], Rackspace, IBM and VMware. Cloud Foundry Foundation comprises 33 members and 42 contributing development businesses. Cloud Foundry support languages like Scala, Ruby, Python, PHP, Node.js, Java and platforms like Play 2.x, Lift, Rails, Sinatra, Spring Framework 3.x, 4.x.

AppFog is a PaaS cloud hosting platform offering services like open-source application development, development flexibility and multi-platform support plus provision of multi-languages (e.g. Node, Java, Ruby, .Net, PHP and MySQL). Initially AppFog was only built for PHP, however now expanding technology support for nearly every programming language. The key feature of this platform is the IPhone Integration app that offers capability to manage, monitor and scale apps from anywhere in the world. Moreover, it supports popular frameworks and runtimes. Additional capabilities include easy-to-use scaling, automatic load balancing and many more [28].

CloudBees was established in early 2010. It is a USA based company that calls its contributors as worker Bees. Most of the Bees share a powerful pedigree in constant code integration, fast delivery, app development and open source. In Sep 2014, CloudBees dropped its PaaS, to emphasize on Jenkins continuous integration. "Jenkins is called an open source continuous integration tool developed in Java [29]." Presently, CloudBees established partnership with Pivotal Labs and uses Cloud Foundry as its PaaS [27]. CloudBees is now well established as the Jenkins Enterprise and became a continuous delivery (CD) market leader. CloudBees offer technology based solutions that empower IT industry to show immediate response for software delivery requirements of the business.

V. SUPPORT FOR DEVELOPER PRODUCTIVITY

PaaS ease life of developers, as it eliminates overhead for resource and application management. Application developer responsiveness is one of key aspect in overall application development lifecycle. PaaS technology allows application developers to develop applications with high responsiveness and eliminate secondary hassles. PaaS application development practices have freed up developers and businesses by eradicating monotonous work. Moreover, it offers freedom and ease to the operations of team while testing the application.

From management viewpoint, PaaS helps in optimum utilization of scarce IT resources. This feature ultimately offers low business start-up cost and better application development. It is achieved through a key characteristic of PaaS, which auto-scale application development procedures. That helps to decrease or increase the server capability according to real-time demand of network traffic. Technology providers are required to load-balance requests across several servers and observe the load on every server as well as to spin up new servers as necessary. Presently, almost all PaaS technology providers offer auto-scaling services to some extent. For example, Google App Engine offers PaaS services, which can automatically deal with the most of secondary software development issues. The Google App Engine is built to abstract away the view of individual servers.

GAE automatically develop data stores in specific servers, and then saves client HTTP session (by default) into these data stores. This entire procedure is very transparent for application developers. Another very popular PaaS provider is "Heroku" that offers automatic session sharing through server instances; however Heroku does not offer transparent auto-scaling. Developers have to see the dashboard and add resources to the application as required.

Pricing is a key aspect that often terrifies developers while changing technology platform. The pricing of PaaS technology tools and services is a significant aspect for services providers. The majority of PaaS service providers offer free trials for new PaaS developers to try-out new technology platform. For smaller PHP websites, such free-tiers are outstanding choices [30].

Another significant feature to consider; is the availability of technology support services. Most of PaaS platform offers support services for developers to solutions the technical problems. For example, CloudBees offers the greatest combination of free and paid support services with technically trained support staff [31].

## VI. PaaS Cloud — a SWOT analysis

### A. Strengths

Cost saving: PaaS cloud services offer freedom to clients (developers and businesses) from a heavy burden of expenses for maintenance, updating network servers, data centers management and purchasing proprietary software. The PaaS service providers manage these issues, while developers and organizations simply pay low monthly subscription fee (pay-as-use model).

No licensing: PaaS allows "all-in one" service model that is based upon a specific subscription fee. Such approaches ease business from complexities of expensive and complicated software licenses that require updating and managing regularly.

Better system management: PaaS cloud based development environment minimize the reliance on external consultancy. Now technology providers are dealing with the installations and updating applications. Issues regarding software incompatibility are not any more clients' problem, so there's no need for external IT consultancy to deal and troubleshoot the organization systems.

Mobility: PaaS is turning out to be a one of preferred choice of globally distributed development teams. Data, code, files and projects stored in the cloud can be accessible virtually from anywhere through an internet connection.

Shared resources: PaaS is significantly addressing the better resource management problems. Instead of wasting costly and valuable technology, power and network resources; cloud computing offers more well-organized and affordable ways to utilize computing resources.

Always up to date: The PaaS clients will always be up-to-date with new technology features, development tools and software updates. The service providers will always be offering the latest version of the technology product. By such features, a development business can get a strategic advantage over competitors because of the considerable upgrading of business technology and processes [32].

Higher security: Cloud computing security is a major concern in recent years. However, PaaS cloud services are not security vulnerable as other cloud computing models. The PaaS service provider cares for the maximum security of the client's personal data and projects. A professional external IT security consultant has got this experience so he/she can effectively care for privacy as compared to clients themselves. The risk of losing data is comparatively lower because of a sophisticated backup facility (at technology provider end). It is a main constituent of the integrated service level agreement [32].

Fast implementation: Through PaaS technology, now developers can develop applications and get them to market faster. The traditional software development life cycle has been cut down through eradicating secondary application development tasks.

Worldwide Access: Cloud products can be used worldwide; not only on the computer wherever the software is installed. The PaaS platform runs in the data center of the service provider where all data is centrally stored.

Therefore, the client is capable to connect with PaaS platform and access his/her data from every computer or mobile terminal with internet access across the globe. This makes development much more flexible and also avoids the data redundancy problems.

*B. Weaknesses*

Cloud lock-in: Currently PaaS facing the biggest issue of vendor lock-in. It means that a customer is dependent on a specific vendor only and cannot move to another platform without bearing considerable technology transformation and switching costs. Clients are locked into the same PaaS platform and cannot switch to other platform due to different technology architecture besides working to similar programming language. However, new PaaS technology platforms are trying to mitigate this issue.

Technical Immaturity: From PaaS point of view, every cloud service provider has its own interface, services, methods, tools and costs. The unfolding nature of the PaaS methodology places everything at risk. For example, services could be worsening, technology prices could be changed immediately or the quality of services could be dropped. PaaS is just at its evolutionary stage. It requires lot more time to acquire a certain maturity level. Standardizing bodies are now standardizing PaaS services, like NIST (National Institute of Standards and Technology) is one of such bodies [6].

Privacy and Control: PaaS cloud security is improved considerably in past few years. Service providers are offering very comprehensive security suits for better protection of client intellectual assets. As, it is in the best interest of service providers because improving security leads to better customer satisfaction. Though if a security breach happens, the client would be on risk not the vendor.

Limited flexibility: PaaS technology can't match flexibility of IaaS (Infrastructure-as-a-Service) technology. Different from IaaS, PaaS clients can't necessarily create and delete various VMs (virtual machines) easily.

Legacy systems: From PaaS technology implementation point of view, small and medium-sized companies are more likely to take the advantage of the cloud rather than bigger organizations which may have complex legacy systems. Another big factor is overall transformation cost. A large business needs to go through huge transformations (from managerial to technological aspects) for adopting new technology architecture.

Network Performance: PaaS based development environment is of two types: one in which developer completely develop online; second where you work offline and then update on PaaS server. In both cases, network performance is very critical. Due to slow or not responding internet connections, there can be some performance issues while you are doing significant updates during development process.

*C. Opportunities*

"Just think and PaaS will develop for you"; such terms are getting popular these days, when someone talks about PaaS. PaaS offers high-level programming capability with considerably less complexities. It is one of the highly marketed advantages of PaaS technology these days. The general application development could be more efficient, as it already has built-in infrastructure. Therefore, application maintenance and enhancement has become easier now.

Agility and flexibility: Small and medium size firms are more agile. Therefore, they are certainly able to move to PaaS and start taking advantages of cloud computing services, innovative tools, fast application development and significant cost-saving.

Growth in cloud services: Currently Cloud computing services is one of the hot business areas worldwide. It will continue to evolve by growing competition among both, new competitors and established players. A number of researches and surveys outlined a huge future growth in PaaS market. Further explanation of this point is given in the upcoming 'future growth' section of this paper.

Better service quality: PaaS vendors always offer newest and updated development platform. PaaS clients usually get the latest development tools, upgraded software versions and the latest bug fixes. This could mean a significant strategic benefit over business competitors because the firm is developing application with most modern and improved technology [33].

Low startup cost. When an organization is going to start a new development business, it does not need to buy sophisticated servers, powerful systems and establishing high-tech data centers. While in case of PaaS technology, clients only need is to establish an agreement with PaaS technology provider using pay-as-use model. The most significant advantage offered by PaaS is that client can upgrade technology capabilities as business technology demands increases [33].

*D. Threats*

Bandwidth Bottleneck: PaaS is offered over the internet, private cloud or by an organizational WAN (wide area network). In all situations, there is a need to guarantee that the network connection is significantly reliable, stable and has satisfactory bandwidth to facilitate all operational users. Bandwidth bottleneck is possible in global development, where some countries still don't support high bandwidth for global collaborative development projects.

User attitude and control: After cloud adoption, organizations are required to transfer 'control' of their information and data to third party. For a number of big organizations, the idea of transferring overall control of software, data and hardware to a third party is not a well-liked thought; because all such data hold critical corporate information and significant commercial secrets.

Data Protection: Besides extensive security improvements in cloud technology platforms, the threats of data lose and theft are always there.

Dependency to the provider: If the PaaS technology provider is not able to deliver its services anymore or goes bankrupt for some other reason, in such situations, the clients are at huge risk. It must be ensured that the clients will remain capable to run their services and business operations normally. If not, the entire dependency could lead to the total disaster [33].

Application Requirements: Client applications have to follow PaaS vendor stated specifications. It perhaps would not be worthy enough to port legacy software to a PaaS. PaaS should only be considered for novel application development with state-of-the-art stacks [31].

Runtime Limitations: Sometimes at PaaS platform, the desired framework version is not available at runtime. Also, sometimes other runtime issues may encounter, like user may not be able to install or have the desired libraries and APIs according to the current application requirements. Moreover, a user can also be blocked from carrying out some operating system level calls.

Add-ons Limitations: PaaS providers recognize that maximum applications need a certain kind of caching layer, backing data store, middleware services and messaging requirements. So, number of PaaS providers offer a comprehensive set of add-ons to bridge the gap between traditional and cloud development. However, existing add-ons cannot be enough or unsuitable for some specific application development. Integrating the user-desired add-ons can be hard or sometimes not possible. So user can face limited control while dealing with multi-tenant add-ons.

## VII. KEY STAKEHOLDERS IN PAAS

Cloud computing is transforming the role of traditional stakeholders of technology infrastructure. Now more and more stakeholders are being added in the computing industry. These stakeholders are not limited to clients or providers. Though, enablers and regulators are also becoming part of cloud paradigm, because of new nature of technology and novel delivery model of cloud computing service.

Some of the key stakeholders of cloud computing are outlined by [34]. After detailed and careful analysis, it is outlined that most of cloud stakeholders are also applicable to PaaS technology context. This is due to a similar nature of SaaS, IaaS and PaaS technology.

From PaaS technology context, one of the key stakeholders is PaaS consumers, clients, or users of the service. These clients could be some independent software vendor (ISV) or business. PaaS providers control the PaaS cloud system to offer services to third parties. These service providers are responsible for the entire PaaS infrastructure, service provision and overall maintenance [34].

Another key stakeholder is technology enabler. [34] used the term of "enablers" signify firms, organizations or any individual that sell services or products that enables the adoption, delivery and use of cloud computing. Enabler stakeholders can be consultancy firms that offer service in PaaS adoption and implementation..

Regulators are another type of PaaS technology stakeholders. Those are one of the key participants of the whole cloud technology infrastructure. Regulators are those who formulate laws and regulations that narrate the cloud computing business. The European Union has guidelines that entail a corporate to know where the data is stored and information about its possession. This is in clash with the normal cloud computing nature where the hosted data of client could reside at any site that the client organization might be unaware of [34].

Network providers, internet service providers, broadband operators and mobile network operators are now also turning out to be one of the key stakeholders for PaaS technology. The idea of incorporating these stakeholders into PaaS paradigm is purely based on provision of the core technology resource and capabilities. Approximately the entire cloud computing services are reliant on the constant availability of the Internet connection, so network providers are one of the key members of the PaaS cloud family [8, 35].

## VIII. PaaS Application Development Life Cycle

How PaaS technology works? What is an application development life cycle in a PaaS cloud development environment? This section is going to answer such questions. PaaS is a set of services that abstract-away the operating system, application development infrastructure, configuration details and middleware. It offers app development teams, the capability to design, develop, test, and deploy applications. PaaS enables software deployment by means of on-demand tools, automation, self-service, state-of-the-art technology resources and a hosted platform runtime container. This eradicates the requirement for a comprehensive installation kit. In addition, software developers are no longer required to configure and wait for virtual machines (VMs) or physical servers to copy files from one platform to another as they transfer over the application development life cycle. PaaS infrastructure modernizes life-cycle supervision, from initial development to final application deployment. PaaS technology automating several phases and functionalities linked with each application development stage milestone. The new modernized web based development environment also simplifies the application patching, version updates and other maintenance tasks.

Fig. 3 shows a PaaS application development life cycle, where PaaS drives an application to the cloud from a command-line infrastructure or straight from an IDE using plug-ins. Next, it analyzes the on-hand applications. Then it starts hosting apps in runtime platform container that analyzes its main resource needs. Along with offering scaling facility, PaaS cloud also offers high automatic configuration, availability, load balancing and management tools. The life cycle of application development at PaaS starts with app build. User can select from many available design templates. In the second stage that is "land first release", users are offered self-service deployment and multitenancy. In the next stage, the user has to maintain application. The last stage is "main land release", where application is ready for delivery.

PaaS is also able to initiate multiple copies of the similar or various clouds. In some conditions, it might require to isolate app from others. In such situation, the application developer is able to make use of common tools and high quality practices to achieve this feature. Because this development life cycle offers an isolated and more secure environment for development. Through PaaS based development approach, business can unite local resources and data for personalized application development and web services.

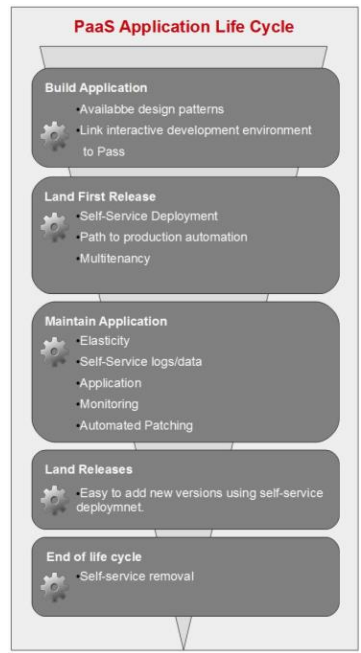

Figure 3 PaaS Application Development Lifecycle

## IX. REVENUE MODELS

PaaS application development life cycle has different phases with different client requirements at every phase. Similarly, PaaS also encompass different revenue models as compared to other technology businesses. PaaS vendors formulated their revenue models based on access patterns of technology resources. In this scenario, pricing signifies the cost which developers spend to join the platform to develop and deploy software. Certain PaaS vendors ask for subscription fee to access for application development, testing services and tools; whereas the other vendors simply charge for the actual time that end-user occupy and use hosted application. For example, the pricing for hosting software with Bungee Connect[1] is calculated through the time that end-user spends interacting with each page of the system. Google App Engine offers a specific amount of technology resources for free and permits developers to purchase extra resources when required. Through analyzing maximum daily cost, the developers are able to allocate quotas for a set of adjustable computing resources. For example, the network bandwidth utilized, amount of stored data and data sent or received etc. Furthermore, the PaaS vendors may offer an online market for the developer's application commercialization. The vendors usually charge developers a certain percentage of revenue, he/she earn from app. A developer will get his stake from the amount that customers have paid, is actually based on specified *"revenue share rate"*. Developer can place and commercialize his/her app on vendor's platform. Developer will get revenue through selling app online, while vendors will get a certain percentage of that revenue.

PaaS applies "two-sided market model" in which the platform facilitates the supply and demand side. In case of "Information and Communication Technology (ICT)" services, a platform vendor enables an environment for software design, development and deployment. Such platform permits application developers to offer service to the end-customer and acquire revenues from it. However, developer pays to PaaS vendor for using the services. Vendor also gets a certain percentage of revenue paid by app's end customers. Therefore, the platform owner produces value for downstream customers and upstream developers and extracts revenue from both sides.

---

[1] http://www.bungeelabs.com/

## X. RECOMMENDATIONS

### A. Recommendations for Businesses

Which business more suits PaaS? PaaS is future of development industry and we have to know that what kind of development business (size/type) more suits PaaS revolution. This section is aimed to outline recommendations for the different types of organizations, those who could be more benefited from PaaS technology and its capabilities. It is already stated that PaaS is the most attractive solution for small to medium sized organizations to change working practices. While large organizations need to tackle enormous challenges before shifting to the PaaS platform. Most businesses will see the increase in agility and cost reduction; once they initiate deploying the PaaS tools.

Application development businesses often faces a number of situations those require active handling; PaaS technology encompasses all such capabilities to handle all such situations. The below section is outlining some of the key dependencies of technology based businesses and how PaaS turn out to be a savior.

Need less time to market the application: application time to market is one of the major pressures for a development business and project members. The capability to rapidly move an app from concept to development and then to market is only possible through PaaS.

Trouble in aligning application use with stakeholders: PaaS offers enriched solutions to develop a rich application portfolio, which leads to better alignment with technology stakeholders.

Want to reduce huge operational costs: The considerable reduction in administration cost for a set of applications, offer a significant influence on the fundamental application development practices. PaaS is a smart choice for organizations, who are seeking less administrative hassle.

Require better project management. In situations, when the loss of a particular administrator has disastrous outcomes for a development project; it may be the right time to converge to PaaS platform. PaaS offers repeatable and formalized procedures for administration tasks and number of centralized management tools. Such procedures lead to better project management. In case, if a manager leaves the organization, still project will be managed and have fewer chances to breakdown because of PaaS based management capabilities.

Need extensive over-provisioned technology infrastructure. If business needs huge server power or high-tech data centers for initiating project, in such situations, PaaS is the best option for initiating such project with a very small initiation cost.

*1) Organizational Implimentation of PaaS*

Adopting PaaS is an extremely complex decision that any organization can make. This section is aimed to offer a list of different aspects, those needed to be considered before planning to adopt PaaS technology services. These are six key recommendations which are essential to be assessed prior to PaaS adoption:

1. Learn more about PaaS. Study technical material and know more and more about PaaS technology. Analyze trends and technology aspects which can facilitate business in near future.

2. Consultation with developer organization to analyze readiness and interest. Through this consultation, management will know about possible benefits and bottle-necks in the process of PaaS adoption.

3. Learning more about PaaS vendor's offerings. Compare and contrast different vendor's services, tools, data rates and middleware. There are many things that are needed to be ensured. For example, multiple programming languages support, data manipulation technologies and application services and most importantly no vendors lock-in.

4. Outline what kinds of applications are desired. Currently, PaaS technology is supporting numerous application architectures. Through analyzing business technology needs, skill set and project requirements, clients are required to make the decision about target application. Below different types of applications are mentioned [36]:
- Service-oriented applications
- Hybrid applications
- Mobile applications

5. Select if want to port existing applications or going to build whole new system. This is one of the crucial decisions that need to take.

6. After assessing and analyzing all above given aspects and areas, there is a need to conduct a proof-of-the-concept.

*B. Recommendations for business professionals*

From the business and technical point of view, when a PaaS technology model is implemented for a business environment, the number of requirements and complexities are prompted. This section is aimed to point out and address the challenges arose from the PaaS infrastructure application.

Private or hybrid cloud: Public PaaS is most common type of PaaS, where business services and infrastructure is hosted by third party. For many organizations, public PaaS appeases to be a little risky. So, another more secure choice is private PaaS, where business services are mostly hosted in an in-house private cloud platform. Private PaaS reside inside the organization and shared by numerous branches or organization departments. Private PaaS is more costly and demands a great deal of investment. A middle ground is hybrid PaaS. Hybrid PaaS have features of both private and public PaaS. It offers partial outsourcing of technology services. All these types of PaaS technology platforms are having different roles, ownership and practice patterns. Individual user or businesses can select the appropriate choice according to business, development and technology needs.

Openness: Many businesses want to integrate the present business services to new PaaS infrastructure. Therefore, it is a major need that PaaS vendor offer openness capability. In this way, PaaS services or solutions should be open for integration with an existing (external or internal) system. This will offer a more adoptable technology transformation for current stakeholder of business.

Industry specific standards: Business industry supports specific industry related guidelines and standards. PaaS technology adoption and application also need to consider and implement such technology based standards for better performance.

Diversity of business model and roles: every business has specific user groups. These user groups have different system usage and user behavior patterns. The PaaS solution should recognize and satisfy such patterns and needs.

Time to deliver the solutions: In the application development industry almost every enterprise is required to rapidly deliver solutions to market. In this scenario, PaaS technology implementation offers quick solutions and don't hinder corporate for long. The reason is that cloud environment offers innovative development lifecycle for better continuity of business.

User: Majority of corporate users do not possess extraordinary programming skills. In this scenario, PaaS offers a great deal of simplified development capability that even support to non-IT professionals to use services to develop, implement and deal with applications.

Business Agility: Specific commercial requests are pushed from business customers/stakeholders to developer. For example, cutting down the project time, change management, support of corporate agility and rapid scaling of the business system etc. Fortunately, PaaS is here for dealing and tackling all these issues and aspects.

*C. Recommendation for developers*

PaaS offers a virtual platform intended for software development and deployment. The user develops software through interacting with the PaaS vendor' servers, normally through a web browser. Software programming model permits developer-specific business process flow and application logic without referring to the fundamental physical network interfaces and computer systems. The model accumulates program logic into the system service calls applied through the client, servers and network infrastructure. PaaS therefore conceals the complexity of logic operating among a client and a virtual server throughout the virtualized infrastructure.

Mobility of data is now one of the key aspects in this new technology era. Whether, users are in office, home or travelling, one can access his/her data. The same idea is valid for PaaS technology cloud. A developer can access applications remotely; he/she has the freedom to access it from anywhere. Cloud based systems store the data online, eliminating the requirement to call data from an outside source and thus minimize database-integration issues.

Developers often concern about resource management. PaaS technology suppliers offer numerous tools that facilitate at different levels of precision for letting business to track the usage of their PaaS resources. They can track aspects for example: application usage details, time spend on a particular app, user tasks, problems appeared, data usage and software's performance etc.

PaaS vendors often try to make the programming experience familiar to developers through offering support to numerous common programming languages. Though, some PaaS vendors like Salesforce.com have developed their own programming languages. They state that new programming languages ease the development procedure and are comparatively easy to learn.

*D. General recommendations*

This section of recommendation is based on more general references to PaaS and cloud computing services. Questions like, "how PaaS technology improve, enhance and revolutionize traditional application development and business practices" would be answered here.

Embrace cloud, both internally and externally: Cloud offers a wealth of technology resources to facilitate business. There is time and cost related profits that a midsize or a small company will see almost instantly (for example, the capability to access technology recourse at a low (pay-as-use) fee, versus making huge upfront capital investments for business initiation). As a business becomes cloud adopted, further gains will turn out to be more apparent, for example, significantly enhanced corporate flexibility.

Tap into the power of web APIs: There is a huge wealth of online technology resources (updated and tested) that can be incorporated into specific operations through a single click. APIs (application programming interfaces) were appeared as technical glue that facilitates to put together applications. Currently, APIs bring businesses together. There is no need to reinvent the wheel when there is already an API waiting to be used.

Build a culture of security: The online world seems to be a very scary place, especially with news about everyday hacks, data theft and frauds. However, cloud vendors have significantly resolved security issues and now PaaS cloud can be perceived as one of the secure technology platform for a development industry.

Discover the power of mobile: Growing number of corporate transactions is happening through mobile technology. Mobile devices likewise offer flexibility and improve employee efficiency. In case of PaaS cloud infrastructure, smart technology of mobile offers employees, customers and partners to be engaged and collaborate while they are out of office or traveling anywhere.

Embrace social cooperation both internally and externally: Association with clients, partners and staff promote innovation and develop brand loyalty. Social cooperation offers a new approach to engage with both employees and clients. For developing firms, mutual communication and coordination is one of the best ways to improve performance and capability.

Explore new business models: PaaS technology revolutionizing development and business capabilities. Transformation of digital business open up opportunities for new services that can be speedily designed and developed as desired. New business models are developing. So, businesses can take advantage through adoption and implementation of these models.

*E. Top Market vendor*

An analysis is performed among top key players of PaaS market. The companies included are Heroku, Amazon, Windows Azure, AppFog, OpenShift, Engine Yard, Google App Engine, CloudBees and Cloud Foundry. This overall analysis was performed on the basis of major services and support capabilities offered by these vendors. A comprehensive analysis is outlined in Appendix A. The analysis reports that current PaaS market is dominated by few top players. Heroku, Amazon, Windows Azure, Google App Engine and Cloud Foundry are top players in PaaS market presently (respectively) as shown in Fig. 4 below. These PaaS vendors are offering excellent services and top capabilities to satisfy the customers' demands.

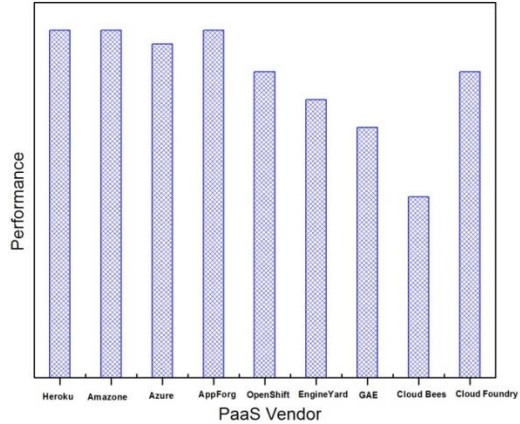

**Figure 4 PaaS Top Market vendors**

## XI. PaaS Future Directions

### A. Novel PaaS Layered Architecture

According to "Gartner's PaaS Road Map report", novel PaaS architecture is divided into 3 following layers:

Layer 1- Application platform as a service (aPaaS): this layer offers a comprehensive application platform. It is utilized by the actual application's constituents (business process supporting components) or through its APIs. Business level users and developers gain speed-to-market, through fast application development capability. Also get the competence to focus on specific application development functionality instead of developing the entire application platform.

Layer 2- Software infrastructure as a service (SIaaS): this layer offers service management for software components. It offers services like online cloud database, messaging and integration. This layer is analogous to the earlier layer, as it offers the development tools to develop software in the cloud. However, it is targeted at developers instead of business level clients.

Layer 3- Cloud enabled application Platform (CEAP): This layer act as software middleware that uphold the private and public cloud features, including complexity management, monitoring, optimization and scaling.

### B. PaaS Core Architecture Transformation

The cloud computing architecture depends on virtualization techniques to accomplish elasticity and productivity for large-scale resource sharing. Till last few years, virtual machines (VMs) were denoted as back bone to achieve virtualization in cloud. However, the idea of containers based OS-level virtualization is changing cloud market significantly.

Containers offer two key benefits. They offer improved performance as compared to VMs based virtualization. Containers facilitate to write apps in scripting languages, that looks more similar to those written in Java and .NET as far as PaaS goes.

Why use Containers? It is an important to know while discussing architectural transformation in PaaS. The virtual machines take up plenty of system resources. Every virtual machine runs not only a whole copy of an OS; it runs a virtual copy of everything (all system resources). This rapidly adds up the huge burden to CPU and RAM cycles. On the other hand, containers requirements are small therefore they enable faster start-up with improved performance, less isolation and superior compatibility.

From PaaS point of view, containers can empower company to pack additional applications into a single physical server as compared to VM. So, container technologies such as Docker have beaten Virtual machines at this part of the data-center game and cloud.

A Docker container is a complete package that comprises of all technology resources that it requires to run like: system tools, code, system libraries and runtime. Docker container can hold anything that can be installed on a server. This assures that it will continually run the same irrespective of the running environment.

Docker open source containers emerged over the past few years and developed a de-facto standard. This container technology offered capability to applications to run as extensions form one platform to another. For example, applications are running as micro-services in Open-Shift and Cloud Foundry PaaS environments. Docker containers have currently became available with all major Linux distributions as well as supported in major cloud services.

Currently every major cloud vendor and enterprise infrastructure software company has jumped on the Docker trend including IBM Corp., Google Inc., Microsoft, RedHat Inc., VMware Inc., and Rackspace. RedHat is presently the top outside open source contributor for containers development and all latest distribution of Red Hat Enterprise Linux offering Docker containers.

Docker containers are playing an important role in transformation of PaaS industry. Now, Docker and related lightweight containers intended to revolutionize the role of operating system and the VMs. It is much like the same as VM has done to the physical bare-metal server infrastructure. Docker containers are getting popular in PaaS industry because they offer less overhead and better interoperability as compared to traditional VMs.

## XII. FUTURE MARKET GROWTH

According to Gartner's report "PaaS Road Map", it is stated that PaaS cloud-based application solutions will grow at a faster speed as compared to on-premises solutions. It is projected that up-to 2015, 50 percent of all ISVs will become SaaS providers. It is expected that PaaS and SaaS technology will become a main part of corporate applications and functionalities directly or indirectly. Fig. 5 shows the PaaS road map till 2015, where it demonstrates past and present innovations and transformations in technology.

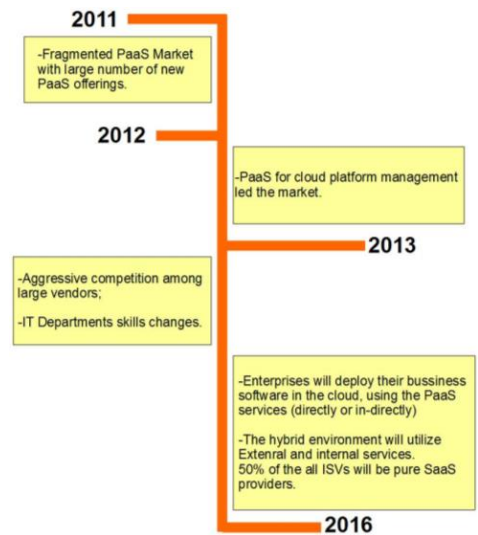

**Figure 5 PaaS Road Map**

At Gartner Application Architecture, Development & Integration Summit 2011 outlined some of the noticeable figures regarding the PaaS technology future. According to report, it was stated that till 2015, enterprise adoption of PaaS will rise from 3% to 43% [37, 38]. According to Visiongain forecast for the PaaS submarket revenue is valued at $1.9 Billion in the year 2013 and it is expected to increase up-to $3.7 Billion till year 2018. Fig. 6 demonstrates the Visiongain's forecast for PaaS till 2018.

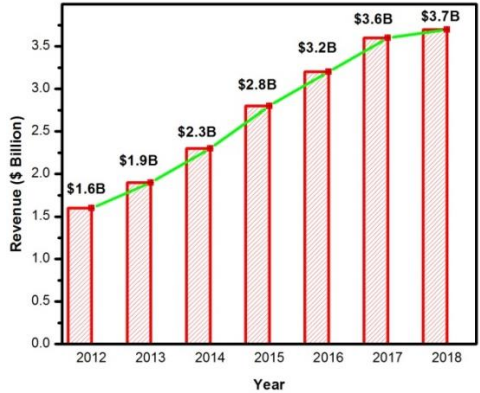

**Figure 6 PaaS Market forecast 2013-2018**

According to the 451-Research analysis report on PaaS future market; it is stated that PaaS will attain 41% CAGR through 2016 by producing 24% of overall cloud computing services revenue. It is also assessed that 71% of PaaS revenues will be produced through vendors over $75M in sales. Another 451-Research report on "The future of cloud services" forecasted that PaaS is a fastest growing industry of the future, as shown in Fig. 7. Hosted Infrastructure Services (69%) and SaaS (71%) are presently two popular cloud services, with the future forecasted growth rate of 14% by 2016. While the fastest growing industry is PaaS, with current 37% growth rate; projected to grow another 26% till 2016.

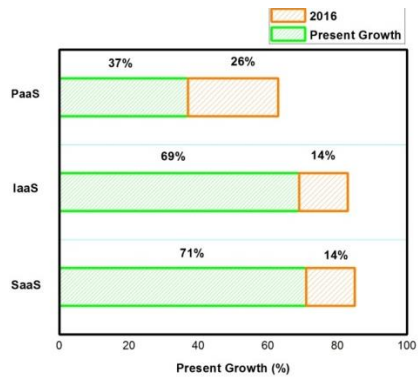

**Figure 7 PaaS, IaaS & SaaS future growth analysis (Current-2016)**

### XIII. CONCLUSION

Cloud computing is our future. Cloud computing is turning out to be back bone of technology industry. Platform-as-a-Service (PaaS) is a revolutionary paradigm of cloud computing, that emerged to offer easy, low cost and agile application development. This paper outlines the business aspects, trends and evolution of Platform-as-a-Service (PaaS) technology. This research is significant because it has offered a deep insight into the changing cloud market trends and future cloud technology hype. Year 2015 is declared as the year of PaaS, so there is a great deal of need to assess the present market situation and analyze how PaaS cloud technology is going to perform in next few years. This research highlights the technology business, developers and common-user aspects for PaaS technology. These aspects include technology platform, basic architecture, business model, revenue model, market forecasts, top technology providers and possible prospects of PaaS cloud computing technology. This paper has outlined key benefits of PaaS technology and performed a SWOT analysis to analyze the technology feasibility. One of the key parts of this research is recommendations, where author has tried to offer some recommendations from PaaS business, developer and user point of view. Analysis among top players of PaaS market outlined the best service providers in current PaaS market. The basic idea is to find out the top PaaS vendor with excellent technology services.

From virtual machine generation of PaaS to new containers based PaaS, a comprehensive new horizon of PaaS technology is emerging and transforming the market and application development practices. This research addresses all these areas in deep and detailed way. PaaS technology is transforming traditional legacy development approaches and offering a new more agile way of application development. These new ways and services are more beneficial, time efficient and economical for businesses. Therefore, the fundamental objective of this paper is to highlight these practices and recommend better development PaaS approach. This research has offered a great deal of insight for developers and businesses to assess the PaaS from different business perspective.

XV. APPENDIX

A. *Appendix A: PaaS Vendors Analysis*
  1) *Key Features*

This category includes key features those are necessary for PaaS development platforms. The table 1 shows a comparison of top PaaS providers. Here it is assessed that Heroku, Amazon, Windows Azure and AppFog are top platforms with maximum features availability for clients.

**Table 1 Key Features by PaaS vensors**

| Features | Redeployment Speed Score | App Monitoring/Analytics | Command line Interface (CLI) | SSL Endpoints | Scaling Web/Worker Roles | Custom Role Definitions | Service Add-ons | Content delivery Networks (CDN) |
|---|---|---|---|---|---|---|---|---|
| **Heroku** | Yes | Yes | Yes | Yes | Yes | Yes | Yes | No |
| **Amazon** | Yes | Yes | Yes | Yes | Yes | No | Yes | Yes |
| **Windows Azure** | Yes | Yes | Yes | Yes | Yes | No | No | Yes |
| **AppFog** | Yes | Yes | Yes | Yes | No | Yes | Yes | No |
| **OpenShift** | Yes | Yes | Yes | Yes | Yes | Yes | No | No |
| **Engine Yard** | Yes | Yes | Yes | Yes | Yes | No | Yes | No |
| **Google App Engine** | Yes | Yes | Yes | Yes | Yes | No | No | Yes |
| **CloudBees** | Yes | Yes | Yes | Yes | No | No | Yes | No |
| **Cloud Foundry** | Yes | Yes | Yes | No | Yes | No | Yes | No |

*2) Frameworks Support*

The table 2 shows a comparison among top PaaS vendors for offering programming language framework support. This comparison shows that Amazon is top PaaS vendor with maximum framework support, while Heroku, OpenShift and AppFog are runners up.

**Table 2 Frameworks Support offered by PaaS vendors**

| Frameworks | PHP | Python | Ruby | ASP.NET | Java | Django | Rails | Node.js | Word Press | Drupal | Joomla |
|---|---|---|---|---|---|---|---|---|---|---|---|
| **Heroku** | Yes | Yes | Yes | No | Yes | Yes | Yes | Yes | Yes | Yes | Yes |
| **Amazon** | Yes | Yes | Yes | Yes | Yes | Yes | Yes | Yes | Yes | Yes | Yes |
| **Windows Azure** | Yes | Yes | No | Yes | Yes | Yes | No | Yes | Yes | Yes | Yes |
| **AppFog** | Yes | Yes | Yes | No | Yes | Yes | Yes | Yes | Yes | Yes | Yes |
| **OpenShift** | Yes | Yes | Yes | No | Yes | Yes | Yes | Yes | Yes | Yes | Yes |
| **Engine Yard** | Yes | No | Yes | No | No | No | Yes | Yes | Yes | Yes | Yes |
| **Google App Engine** | Yes | Yes | No | No | Yes | Yes | No | No | No | No | No |
| **CloudBees** | No | No | No | No | Yes | No | No | No | No | No | No |
| **Cloud Foundry** | Yes | Yes | Yes | No | Yes | No | Yes | Yes | No | No | No |

*3) Database Options*

In this category, table 3 shows database capabilities offered by different PaaS vendors. It can be assessed that most of PaaS providers are performing well. The notable thing in this comparison is that big names like Amazon and Engine Yard are lacking behind in this battle.

**Table 3 PaaS Vender Database Options**

| Database Options | PostgreSQL | MySQL | MongoDB/NoSQL | Blobs/Binary Storage |
|---|---|---|---|---|
| **Heroku** | Yes | No | Yes | Yes |
| **Amazon** | No | Yes | Yes | No |
| **Windows Azure** | No | Yes | Yes | Yes |
| **AppFog** | Yes | Yes | Yes | No |
| **OpenShift** | Yes | Yes | Yes | No |
| **Engine Yard** | Yes | Yes | No | No |
| **Google App Engine** | No | Yes | Yes | Yes |
| **CloudBees** | Yes | Yes | Yes | No |
| **Cloud Foundry** | Yes | Yes | Yes | No |

*4) Deployment Options*

Table 4 shows a comparison of different development options like Command-line Interface, Git and FTP. In this analysis it is found that Windows Azure and AppFog are top of list with wide ranging development support capability. Another noticeable thing in this comparison is that Google App Engine is way behind as compared to other top PaaS vendors.

**Table 4 Deployment Options Supported by PaaS Vendors**

| Deployment Options | Command-line Interface | Git | FTP |
|---|---|---|---|
| **Heroku** | Yes | Yes | No |
| **Amazon** | Yes | Yes | No |
| **Windows Azure** | Yes | Yes | Yes |
| **AppFog** | Yes | Yes | Yes |
| **OpenShift** | Yes | Yes | No |
| **Engine Yard** | Yes | Yes | No |
| **Google App Engine** | Yes | No | No |
| **CloudBees** | Yes | Yes | No |
| **Cloud Foundry** | Yes | Yes | No |

*5) Support*

Without technical support a businesses or services are incomplete. So, table 5 shows a comparison of top PaaS vendors for offering technical support. In this scenario, it is assessed that Google App Engine is in top position with maximum offerings of technical support capability for the clients. While big vendors like OpenShift need more improvements in this category.

**Table 5 Technical Support offered by PaaS vendors**

| Support | Technical Documentation | FAQs | Ticket-based Support | Email | Phone |
|---|---|---|---|---|---|
| **Heroku** | Yes | Yes | Yes | Yes | Yes |
| **Amazon** | Yes | Yes | Yes | Yes | Yes |
| **Windows Azure** | Yes | Yes | No | Yes | Yes |
| **AppFog** | Yes | Yes | Yes | Yes | No |
| **OpenShift** | Yes | Yes | No | No | No |
| **Engine Yard** | Yes | Yes | Yes | No | Yes |
| **Google App Engine** | Yes | Yes | Yes | Yes | Yes |
| **CloudBees** | Yes | Yes | Yes | No | No |
| **Cloud Foundry** | Yes | Yes | Yes | Yes | Yes |